Nyagudi Musandu Nyagudi, Independent Researcher and Security Analyst, Nairobi, KENYA
ORCID ID : http://orcid.org/0000-0001-5809-2209
Google Citations : https://scholar.google.com/citations?user=YURdCNoAAAAJ&hl=en
Contact : nyagudim@yahoo.com

# Clinical Informatics : Recursive Concurrence Intravenous Medication Administration Systems Protocols for Addressing the Potential Problem of Recessive Lethal Autonomy in Smart Infusion Systems [12]

NYAGUDI Musandu Nyagudi,

B.Sc. Math & Phy (Nairobi, 1997), M.Sc. Info. Sys. (Notarized Equiv. of Nairobi, 2015),

Cert. Net. Supt. Spc. (IAT Kenya, 2007),

Researcher and Peer Reviewer on The Informatics of Lethal Autonomy

Nairobi, KENYA.

*Author reports that he has no conflict of interest issues in relation to the subject matter.*

## Abstract

Smart Infusion Pumps are vital tools for use in administering a broad range of parenteral/intravenous medications. Safe and effective use of smart infusion pumps depends upon their integration with Pump Servers, Computerized Physician Order Entry Systems, Pharmacy Information Systems, DERS/MERS – Dose/Medication Error Reduction Systems( and the digital Drug Libraries that they use), eMARS – electronic Medication Administration Records Systems, etc. More computer systems

**Nyagudi Musandu Nyagudi, Independent Researcher and Security Analyst, Nairobi, KENYA**
**ORCID ID : http://orcid.org/0000-0001-5809-2209**
**Google Citations : https://scholar.google.com/citations?user=YURdCNoAAAAJ&hl=en**
**Contact : nyagudim@yahoo.com**

result in more computer network connections, with systems that run on ever more lines of computer software code. Resulting in more chaos and more complexity, and inevitably a greater likelihood for systems failure and/or malfunction - this in effect runs contrary to the objective of deploying clinical informatics solutions such as smart infusion systems, due to their inherent technical and chronic performance problems as well as error prone usage that may occasion Recessive(unintended but underlying) Lethal Autonomy.

The Recursive Concurrence protocols suggested in Nyagudi[12] are a means for reducing if not eliminating the challenging phenomenon of Recessive Lethal Autonomy in clinical informatics devices and/or systems.

## Objective

To analytically appraise a broad range of research papers as the primary dataset, and

**Nyagudi Musandu Nyagudi, Independent Researcher and Security Analyst, Nairobi, KENYA**
**ORCID ID : http://orcid.org/0000-0001-5809-2209**
**Google Citations : https://scholar.google.com/citations?user=YURdCNoAAAAJ&hl=en**
**Contact : nyagudim@yahoo.com**

determine if there is scope for occurrence of Recessive Lethal Autonomy affecting human patients, being treated with smart infusion pump systems. This resultant paper would enable practitioners and manufacturers to limit or eliminate realistic problems even if they are only conceptualized as hypothetical.

## Open Access Sources

The research papers analyzed are obtained from Open Sources over the Internet, ie. Open Access type digital libraries via Google and Google Scholar search engines. A simple search run with the title and author(s) names should return one or more uniform resource locators of the references.

## Criteria for Use in Analysis

Medical Informatics papers selected and analyzed contain the required subject-matter and are very recent in most instances. Analysis of these reports/papers was the basis for establishing substantial plausibility of the research topic.

## Methodology

A list of relevant papers was generated. Assessments/analysis for potential Lethal

**Nyagudi Musandu Nyagudi, Independent Researcher and Security Analyst, Nairobi, KENYA**
**ORCID ID : http://orcid.org/0000-0001-5809-2209**
**Google Citations : https://scholar.google.com/citations?user=YURdCNoAAAAJ&hl=en**
**Contact : nyagudim@yahoo.com**

Autonomy issues was undertaken with the recognition that the ever growing complexity of Smart Infusion Pump Systems offers not only the presumed enhanced safety and efficacy, but also the occasional slip through of Recessive Lethal Autonomy and the resultant patient death or injury. Lethal Autonomous procedures would not be intentionally installed in medical smart infusion pump systems – this analysis does not cover the network cyber-security aspect of such possibilities.

**Analysis**

Apart from skill-sets in nursing, pharmacy and medicine, one ought to develop a well grounded understanding of Informatics to grasp the complexity of errors, misconceptions and defects that may occur during the use of sophisticated computer controlled medical systems e.g. Smart Infusion Pump Systems.

**Discussion : The Problem of Recessive Lethal Autonomy in Smart Infusion Systems and suggested Solutions**

*Use of Integrated Smart Infusion Systems in Treating Body Part "X" Malignant Lesions.*

**Nyagudi Musandu Nyagudi, Independent Researcher and Security Analyst, Nairobi, KENYA**
**ORCID ID : http://orcid.org/0000-0001-5809-2209**
**Google Citations  : https://scholar.google.com/citations?user=YURdCNoAAAAJ&hl=en**
**Contact : nyagudim@yahoo.com**

This hypothetical example is given to illustrate the nature of complexities that smart infusion systems are yet to address/resolve.  An integrated  smart infusion system would include eMARs( electronic medication administration records), Pharmacy Information Systems, DERS/MERS(Dose/Medication Error Reduction System), CPOE(Computerized Physician Order Entry) and updated Drug Libraries.  But the smart/intelligent functionality of these systems is pretty limited.

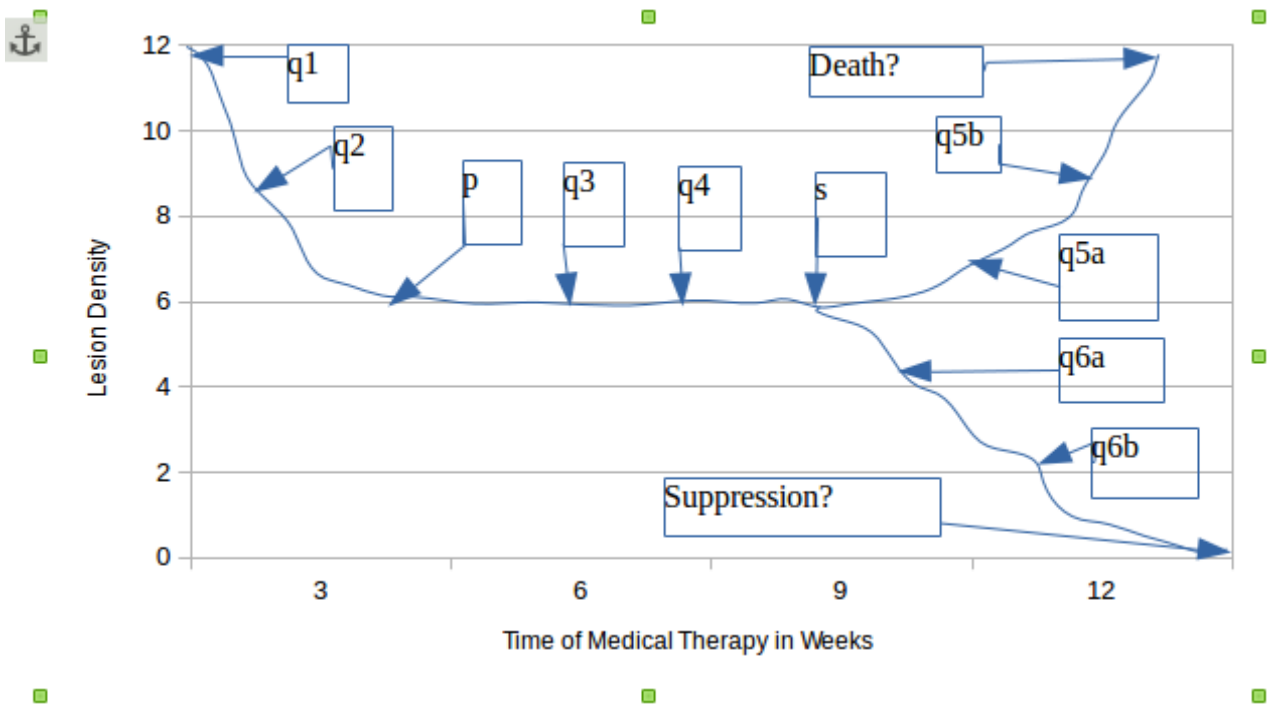

GRAPH 1 is purely hypothetical and analytic in nature, it illustrates the relationship between the parameters of Body Part "X" Malignant Lesion Density vs. Time Period of Medical Therapy with an integrated smart infusion pump system.

At point q1 the medical therapy begins with the use of a smart infusion system.

At point q2 during the continuation of the medical therapy the patient's prognosis is that of a receding density of lesions as determined by way of imagery scans and blood works, as well as general disposition of the patient.

From points p to s diagnostics tests are done at times q3 and q4 and the patient's condition seems to have stabilized. So slightly before time s onwards, the wrong or right decisions may be :

*Wrong Decision* : Medical Therapy is terminated or substituted with a less effective type before or at time s. Consequently tests at points q5a and q5b indicate a condition of increasing lesion density and the patient eventually passes on.

*Right Decision* : Medical therapy continues or is substituted with a more effective type. Consequently tests at point q6a and q6b return the prognosis of an improving patient, to the point of healing or maximum possible sustained suppression of the disease.  This would allow for much lower and sustained dosages to be introduced in the long run to ensure there is no major recurrence of the lesions.  Smart infusion systems are yet to be equipped with sufficient levels of Artificial Intelligence for them to read and understand the curve in GRAPH 1.

A smart infusion system would not detect the molecular oncology implications of its functions, e.g. biopsy or scan results.  It is currently smart only to the extent that it would administer the correct therapy, at the correct dose at the correct rate and time, to the correct patient, if it is well managed.  But at no time would it perform complex and independent analysis of the patient and demand/request for more information or make a medical decision or provide decision support thereafter.

A broad spectrum of Artificial Intelligence Systems would be required to make smart infusion pump systems, practically smart e.g. Machine Learning to assess blood works results, image processing to deduce lesion density from imaging scans, case based systems to call for more tests or to render decisions .  Lesion density could be determined by taking non-intrusive internal imagery scans from different angles and evaluating colour contrast.

**Nyagudi Musandu Nyagudi, Independent Researcher and Security Analyst, Nairobi, KENYA**
**ORCID ID : http://orcid.org/0000-0001-5809-2209**
**Google Citations : https://scholar.google.com/citations?user=YURdCNoAAAAJ&hl=en**
**Contact : nyagudim@yahoo.com**

In current medical practice diagnostic lab technicians, oncologists, radiologists, general practitioners, nurses and other medical/nursing care givers do not meet at one instance to analyze a specific patient's diverse sets of diagnostic data, hence the lack of direct co-ordinated collaboration that may result in wrong diagnosis. An intelligent smart infusion system would fill in this void, resulting from lack of direct collaboration, where Medical Specialists cannot fully understand diagnostic data and render appropriate interpretations due to workload and work environment issues.

Intelligent Architectures for smart infusion systems would facilitate the development of medical context-rich and molecular detail specific Big Data Analytics for cancer, anesthesia, etc. resulting in newer more effective medical procedures.

### Smart Infusion Pump Safety Measures

Smart Infusion Pump systems come with various levels of safety measures in-built. The most notable is the drug library, that determines the hard(unadjustable) and soft(possible to override) limits for medication administration. The drug library is utilized by Dose/Medication Error Reduction Systems(DERS/MERS) to issue alerts/alarms when these limits occur.

**Nyagudi Musandu Nyagudi, Independent Researcher and Security Analyst, Nairobi, KENYA**
**ORCID ID : http://orcid.org/0000-0001-5809-2209**
**Google Citations : https://scholar.google.com/citations?user=YURdCNoAAAAJ&hl=en**
**Contact : nyagudim@yahoo.com**

Drug libraries are the most critical safety measure in existing smart infusion systems, they are frequently updated automatically via the Internet or manually in accordance with hospital pharmacy policy. Poor calibration of drug library limits may result in unintended safety alerts or no safety alerts at all, with great risk of potential harm to the patient's health/well-being.

Alarm fatigue may occur due to frequent unintended alarms as a result of poor calibration of drug library limits, the consequence to patient safety would be that nursing staff may become tired of reviewing alerts, enhancing the risk of a critical alert being overlooked and a patient being harmed during parenteral therapy.

Human-on-the-Loop functionality in responding to smart infusion pump alerts is a major safety provision in these systems. But this can only prevent Recessive Lethal Autonomy to the extent that the Drug Library limits are correct and the computer source code for the DERS/MERS engine that is reading the limits and actuating the pump functions has no errors.

A casual check on the Internet  at the time of research for this analytic paper

**Nyagudi Musandu Nyagudi, Independent Researcher and Security Analyst, Nairobi, KENYA**
**ORCID ID : http://orcid.org/0000-0001-5809-2209**
**Google Citations : https://scholar.google.com/citations?user=YURdCNoAAAAJ&hl=en**
**Contact : nyagudim@yahoo.com**

demonstrates that smart infusion pumps software in many cases have proprietary and unknown code, that is not open to scrutiny. This is contrary to the norms of Nursing, Pharmacy and Medical practice, where all methods are open to scrutiny.

In Mansfield, J., & Jarrett, S.[8] the possibility of System Error is laid out by citing the report "To Err Is Human : Building a Safer Health System", Committee on Quality of Health Care in America, The National Academies Press, 2006 – to the extent that medical errors when using equipment such as smart infusion pumps are predominantly the result of system errors, rather than the error of human operators.

Pharmacy, Medicine and Nursing practice regulatory bodies have not put forward the demand for open-source code disclosures of all software used in critical clinical systems in the same way that they are the custodians and evaluators of essential formula and procedure in the realm of health care. This may be due to the fact that computer programming is not a core competence in these fields at least for the moment.

The complexity of networked control/computing systems and smart infusion systems

**Nyagudi Musandu Nyagudi, Independent Researcher and Security Analyst, Nairobi, KENYA**
**ORCID ID : http://orcid.org/0000-0001-5809-2209**
**Google Citations : https://scholar.google.com/citations?user=YURdCNoAAAAJ&hl=en**
**Contact : nyagudim@yahoo.com**

may introduce new unknown types of malfunctions and medical errors. Probably the most lethal possible error shall be by way of some recessive in-built system fault at microscopic/nanoscopic or coding levels, one that is not known to the infusion pump operators and manufacturers. There are unlikely to be any set training schemes and procedures for handling unknown situations. Recessive problems can only be discovered by the system operators as they occur and adversely affect patient safety.

Other problems could be of unintended systems architecture type e.g. the use of radio-frequency identity tags instead of bar-codes, in the switch key function for activating a smart infusion pump via providing clinician identity and patient identity details. This is due to the possible but inadvertent accident sending data from radio-frequency identity tags to a wrong smart infusion pump nearby in the range of reception, something that Trbovich et al[11: 18] did not anticipate.

Suffice it to state that if pharmaceutical standard preparations are open source even when protected by intellectual property laws of patent, it would be difficult to justify why computing/control source codes of smart infusion pump systems are not open source.

Nyagudi Musandu Nyagudi, Independent Researcher and Security Analyst, Nairobi, KENYA
ORCID ID : http://orcid.org/0000-0001-5809-2209
Google Citations : https://scholar.google.com/citations?user=YURdCNoAAAAJ&hl=en
Contact : nyagudim@yahoo.com

A notable challenge is that Mansfield and Jarrett, 2015 was an intra-organzational study work, unless its publication is responded to by the relevant practitioners there is no indication of its immediate adoption.

As smart infusion pumps function they generate alerts/alarms, functions, conditions and responses data, these are also known as Continuous Quality Improvement (CQI) data (Jog et al, 2015)

*Examples of Intravenous Medication Administration Recursive Concurrence Protocols for Addressing Recessive Lethal Autonomy*

Reduction of the probability for occurrence of errors or undetected malfunctions, can be achieved by introducing technically sound and clinically feasible practices that diminish the likelihood of problematic events. For example on keypad entries into the smart infusion system, the likelihood of erroneous entries can be limited by ways of concurrence i.e. a requirement that two people enter data before it is accepted by the system. e.g. probability of one error/malfunction occurring if Operator X uses the keypad of the smart infusion pump or checks alerts/alarms $N(x)$ times is $1/N(x)$

**Nyagudi Musandu Nyagudi, Independent Researcher and Security Analyst, Nairobi, KENYA**
**ORCID ID : http://orcid.org/0000-0001-5809-2209**
**Google Citations : https://scholar.google.com/citations?user=YURdCNoAAAAJ&hl=en**
**Contact : nyagudim@yahoo.com**

where $N(x) \geq = 1$.

The probability of the same for Operator Y is $1/N(y)$ where $N(y) \geq = 1$

If both Operators X and Y must counter-check each others work on the smart infusion pump and approve each others keypad entries by concurrence, inevitably the probability for error/malfunction diminishes to

$1/N(x)$ x $1/N(y) = 1/\{N(x)N(y)\}$ or to much less

This is if we work on the assumption of Operator X is independent of the efficacy of Operator Y and vice versa at using the smart infusion pump system i.e. if all else is held constant.

Therefore the probability of the joint events of concurrence of their operations proves that the chances of fault and/or error occurring under their joint watch and use of the smart infusion systems is greatly diminished

$Pr\{1/N(x) \cap 1/N(y)\}=Pr\{1/N(x)x1/N(y)\}=Pr\{1/\{N(x)N(y)\}\}$......................[19]

$Pr\{1/N(x)\}>Pr\{1/\{N(x)N(y)\}\}$...................................................................[19]

$Pr\{1/N(y)\}>Pr\{1/\{N(x)N(y)\}\}$...................................................................[19]

**Nyagudi Musandu Nyagudi, Independent Researcher and Security Analyst, Nairobi, KENYA**
**ORCID ID : http://orcid.org/0000-0001-5809-2209**
**Google Citations : https://scholar.google.com/citations?user=YURdCNoAAAAJ&hl=en**
**Contact : nyagudim@yahoo.com**

But for these assumptions to hold true the smart infusion systems would require to take inputs from X and Y independently and make no execution of actuation if there is no concurrence between X's and Y's inputs. Additionally X and Y should not influence each other, i.e. they should make independent decisions.

If smart infusion pump systems were really smart, and endowed with adequate Artificial Intelligence, such systems would make a best effort decision i.e. either aborting the intravenous infusion procedure or making a suggested procedure (i.e. decision support ) available and proceeding in the absence of its approval. Operators X and Y would always have the human-on-the-loop option of aborting all procedures by way of concurrence.

Other issues that may be covered in smart infusion pump systems with enhanced intelligence may be provisioning for autonomous detection of secondary infusion errors, e.g. bag misalignments, secondary clamp maladjustments and tube connectivity/layout. Simple transducer solutions built using existing technologies, maybe engineered to prevent those kinds of problems[11].

**Nyagudi Musandu Nyagudi, Independent Researcher and Security Analyst, Nairobi, KENYA**
**ORCID ID : http://orcid.org/0000-0001-5809-2209**
**Google Citations : https://scholar.google.com/citations?user=YURdCNoAAAAJ&hl=en**
**Contact : nyagudim@yahoo.com**

Trbovich et al 2009 introduces an interesting concept of a confederate nurse monitoring, the smart infusion systems use experiments. This concept could be extrapolated and actualized in a more developed way to enhance smart infusion system safety. There could be a Commanding Nurse and an Executive Nurse at the place of smart infusion medication ( the issues of concurrence of Operators X and Y actualized).

Both Commanding Nurse and Executive Nurse counter-check each other and no action proceeds without their concurrence in intravenous smart infusion medication administration. The smart infusion system would require data entry from both of the Nurses to function, this would reduce errors in many instances.

The Executive Nurse would handle and check the patient while the Commanding Nurse reads the protocols and parameters, and both Nurses counter-check the machine and patient's set-up. The Nurses' obligations/duties in the course of concurrence procedures, would be written into their employment contracts.

An Alternative set-up would be with two Executive Nurses handling the smart

Nyagudi Musandu Nyagudi, Independent Researcher and Security Analyst, Nairobi, KENYA
ORCID ID : http://orcid.org/0000-0001-5809-2209
Google Citations  : https://scholar.google.com/citations?user=YURdCNoAAAAJ&hl=en
Contact : nyagudim@yahoo.com

infusion system and patient, with a Commanding Nurse only reading out the protocols and parameters while counter-checking for errors in the work of Executive Nurses.  There would be a resultant reduction in human errors.

Systems of this suggested design would then produce a soft-copy Electronic "print-out" of the smart infusion system operations to be time-stamped and signed by the Commanding Nurse and Executive Nurse(s).  "Print outs" may assist in "catching" errors if reviewed in a timely manner, especially when Drug Libraries are not used under rare circumstances.

The Novel safety measures  suggested in this section are :

1. Double entry into the input devices by the Executive Nurses (concurrence)

2. A Commanding Nurse reading out  the protocols and procedures to the Executive Nurses and counter-checking the implementation of the same.

3. An Electronic soft-copy "print out" reviewed and signed in the system  by the Nurses involved and also available in hard-copy after the signing.

Smart infusion pumps should warn the Nurses when there are too many overrides,

further more a concurrence approach would greatly limit the possibility of unchecked overrides, and act as a catalyst for future planning. The presence of two or more empowered people counter-checking every action, reaction or alert enhances safety. Logs would also contain the identities of those invoking overrides.

*Tackling Chronic Performance Problems in Control/Computing modules of Smart Infusion Systems*

The vast range of components forming an integrated smart infusion system could malfunction at intrinsic, intranet and/or extranet levels due to their intrinsic software environments and/or hardware platform defects. The resultant chronic performance problems would degrade/compromise smart infusion pump safety, and endanger the patient. Nyagudi[13] suggests a hypothetical technological solution that can be tweaked from the military realm to solve this problem and to guarantee optimal availability of smart infusion system components.

Smart infusion pumps access the most valuable asset in the world, the human blood stream that sustains human life, they must be administered with the most secure

**Nyagudi Musandu Nyagudi, Independent Researcher and Security Analyst, Nairobi, KENYA**
**ORCID ID : http://orcid.org/0000-0001-5809-2209**
**Google Citations : https://scholar.google.com/citations?user=YURdCNoAAAAJ&hl=en**
**Contact : nyagudim@yahoo.com**

system user protocols. Accessing and altering the valuable human blood stream using a helpful but error prone technology, must come with sufficient technological and procedural safeguards.

Concurrence safety measures are used in a wide range of mission critical systems e.g. in military, banking/finance, to reduce or eliminate the likelihood of easily avoidable single-human initiated accidents/errors. A good example of challenges that could be eliminated in this way would be the double bounce inputs on keypads.

Chronic Performance Problem mitigation measures in computing environments such as those of smart infusion systems would require full-time presence and attention of informatics-aware nurses to effect. This would ensure that any complex malfunction that even deactivates alerts/alarms resulting in Recessive Lethal Autonomy are caught at the earliest instance.

## *Big Data Analytics*

The autonomous functioning of smart infusion pumps should not give clinical staff

Nyagudi Musandu Nyagudi, Independent Researcher and Security Analyst, Nairobi, KENYA
ORCID ID : http://orcid.org/0000-0001-5809-2209
Google Citations : https://scholar.google.com/citations?user=YURdCNoAAAAJ&hl=en
Contact : nyagudim@yahoo.com

time to escape from the reality of rare but probable patient harm. The presence of nursing staff in the vicinity of the smart infusion procedure is a small price to pay. Compulsory passive data transfer from smart infusion systems to Analytic systems would further improve the effecting of future safety enhancements after Big Data Analytics[12].

Autonomous remote-query [9:17] is not sufficient for safely loading an obtained Electronic Health Record into a smart infusion system. Due to the comprehensive and complex nature of a well composed Electronic Health Record, use of Artificial Intelligence e.g. Expert Systems and Machine Learning, may be required to reduce an Electronic Health Record into information usable in control of a smart infusion pump.

The deployed Expert System may come up with the conclusion that the Electronic Health Record is incomplete and that an intravenous medication infusion procedure cannot proceed – this would necessitate the intervention of human medical experts, to ensure compliance with the alert after the necessary clinical decisions.

In the absence of some form of Artificial Intelligence reviewing the Electronic Health

**Nyagudi Musandu Nyagudi, Independent Researcher and Security Analyst, Nairobi, KENYA**
ORCID ID : http://orcid.org/0000-0001-5809-2209
Google Citations : https://scholar.google.com/citations?user=YURdCNoAAAAJ&hl=en
Contact : nyagudim@yahoo.com

Record obtained by remote-query by a smart infusion pump, there would be questions in the "World of Informatics" as to the smartness of the system. Artificial Intelligence would also scan Continuous Quality Improvement data during an intravenous infusion as well as sensor data, to render decisions and/or decision support.

Despite the assessments[8] the absence of clinically significant alerts during parenteral therapy with a smart infusion pump, may not necessarily translate to the absence of occurrences/events with clinically significant adverse effects for a patient after a procedure, e.g. failure to recover from anesthesia or any wrong combination of medications that go undetected.

The [7] study describes how Southhampton University Hospitals NHS Trust have developed their own kind of analytics unlike the Purdue University approach known as Infusion Pump Informatics[6] that is a Centralized Repository with computerized analytic tools serving dozens of hospitals, the Southhampton approach in 2009 was a human-centric multi-disciplinary work group.

**Nyagudi Musandu Nyagudi, Independent Researcher and Security Analyst, Nairobi, KENYA**
**ORCID ID : http://orcid.org/0000-0001-5809-2209**
**Google Citations : https://scholar.google.com/citations?user=YURdCNoAAAAJ&hl=en**
**Contact : nyagudim@yahoo.com**

Both the Southhampton and Purdue efforts were launched independently in 2009. The researcher who composed this analytic paper did not find any literature on the convergence of the two projects to date, an occurrence that would be beneficial to the wider Medical Informatics Community.

The Purdue Study [6] effort places emphasis on the post infusion Big Data Analytics of Continuous Quality Improvement data, while the Southhampton effort places emphasis on Drug Library development accuracy. In the quest for the perfect Southhampton Drug Library to be utilized by the Medication/Dose Error Reduction Software, there has been the challenge of conflicting medication administration requirements of different clinical expert specializations. Before extra-clinical area of specialization rationalization of dosages, there is an additional challenge of intra-clinical expert area of specialization rationalization.

**Conclusion**

Artificial Intelligence, Big Data Analytics, Concurrence Protocols, Counter-chronic performance problems protocols, etc. would reduce possible patient harm, in the course of medication administration with use of smart infusion pumps.

**Nyagudi Musandu Nyagudi, Independent Researcher and Security Analyst, Nairobi, KENYA**
**ORCID ID : http://orcid.org/0000-0001-5809-2209**
**Google Citations : https://scholar.google.com/citations?user=YURdCNoAAAAJ&hl=en**
**Contact : nyagudim@yahoo.com**

## References

[1] Jog, Y., Sharma, A., Mhatre, K., & Abhishek, A. (2015). Internet of Things As A

Solution Enabler In Health  Sector. *International Journal of Bio-Science and*

*Bio-Technology 7*(2), 9 – 24

DOI : http://dx.doi.org/10.14257/ijbsbt.2015.7.2.02

[2] Gumpper, K., & Fox, B. (2015). PRACTICE REPORT Pharmacy Informatics

national survey on informatics in U.S. hospitals – 2015. A*MERICAN JOURNAL*

*OF HEALTH-SYSTEM PHARMACY – AJHP : OFFICIAL JOURNAL OF THE*

*AMERICAN SOCIETY OF HEALTH-SYSTEM PHARMACISTS. APRIL 2015*

*Am  J Health-Syst Pharm. Vol.72 Apr. 15, 2015 :* 636 - 55

DOI: 10.2146/ajhp140274

[3] Pedersen, C., Schneider, P., & Scheckelhoff, D. (2015). ASHP national survey of

pharmacy practice in hospital settings : Dispensing and administration – 2014.

*AMERICAN JOURNAL OF HEALTH-SYSTEM PHARMACY – AJHP :*

*OFFICIAL JOURNAL OF THE AMERICAN SOCIETY OF HEALTH-SYSTEM*

*PHARMACISTS. JULY 2015. Am J Health-Syst Pharm. Vol. 72 Jul 1, 2015 :* 1119

– 37

DOI : 10.2146/ajhp150032

[4] Logan, M., Blake, N., Chapman, R., Benjamin, B., Flack, M., Neder, M., Plohal, A., Sims, N., Sparnon, E., Trbovich, P., & Weinger, M. (2015). The Big Picture : A Roundtable Discussion – Working Towards Safer, Easier-to-use Infusion Systems. *Horizon Fall 2015:* 6 – 12. *(* Anesthesia Patient Safety Foundation (APSF) : www.apsf.org American College of Clinical Engineering )

[5] Russell, R., Triscari, D., Murkowski, K., & Scanlon, M. Impact of Computerized Order Entry to Pharmacy Interface on Order-Infusion Pump Discrepancies. : 1 – 21. Milwaukee, Wisconsin; USA : Department of Pediatrics; Division of Critical Care; Medical College of Wisconsin;

[6] Catlin, A., Malloy, W., Arthur, K., Gaston, C., Young, J., Fernando, S., & Fernando, R. (2015). MEDICATION – USE TECHNOLOGY : Comparative analytics of infusion pump data across multiple hospital systems. *AMERICAN JOURNAL OF HEALTH-SYSTEM PHARMACY : OFFICIAL JOURNAL OF THE AMERICAN SOCIETY OF HEALTH-SYSTEM PHARMACISTS. FEBRUARY 2015. Am J Health-Syst Pharm. 2015; 72:* 317- 24 DOI : 10.2146/ajhp140424

[7] Grooves L. (2009).  Innovation & Collaboration : Developing a drug library for "smart" IV infusion devices : a smart move? *Clinical Pharmacist, October 2009, Vol. 1 :* 418 – 21

[8] Mansfield, J., & Jarrett, S. (2015).  Optimizing Smart Pump Technology by Increasing Critical Safety Alerts and Reducing Clinically Insignificant Alerts. *Hospital Pharmacy – Thomas Land Publishers, Inc. www.hospital-pharmacy.com Hosp Pharm 2015; 50:* 113 – 117

DOI : 10.1310/hpj5002-113

[9] Prusch, A., & Small, R. (2015, February).  *A Continuing Education Monograph for Pharmacists : IV Integration and a Culture of Safety : Reducing Complexity and its Consequences :* 1 – 26.  Bartlett, IL :  A Pro CE, Inc.

[10] Simpao, A., & Galvez, J. (2015, June). Current and Emerging Technology in Anesthesia.  *ANESTHESIOLOGY NEWS, ANESTHESIOLOGYNEWS.COM JUNE 2015*: 1 – 8  Mac Mahon Publishing Group.

[11] Trbovich, P., Jeon, J., Easty, A., Fan, M., Pinkey, S., & Rothwell, S. (2009, April). Smart Medication Delivery Systems : Infusion Pumps. *HEALTHCARE HUMAN FACTORS – University Health Network:* 1–98.  Toronto, Ontario, Canada: Healthcare Human Factors Group, Centre for Global eHealth

Innovation

[12] Nyagudi, Musandu (2016). **Clinical Informatics : Recursive Concurrence Intravenous Medication Administration Systems Protocols for Addressing the Problem of Recessive Lethal Autonomy in Smart Infusion Systems.** Nairobi, KENYA : Independent Researcher

[13] Nyagudi, M., (2014). Tackling the "Gremlins Paradox": Autonomous Hot Swap Healing Protocol for Chronic Performance Problem Aversion in On-board Mission Critical Systems on Lethal Autonomous Weapons. Cornell University Library, *arXiv e-Repository – 1409.4409v1.pdf :* 1 – 14

[14] Whitten, J., Bentley, L., & Dittman, K. (2001). *SYSTEMS ANALYSIS AND DESIGN METHODS* (5[th] ed.). New York : McGraw-Hill Higher Education

[15] **,:**enago English Academic Writing – Common Errors in Research Papers, www. enago.com 1 – 19

[16] Public Affairs Directorate, (2014, Michaelmas term). UNIVERSITY OF OXFORD – STYLE GUIDE, Oxford, UK : University of Oxford

[17] University of Victory – School of Nursing (2010). February 4, 7 – 9 p.m. Writing Workshop #6 : 1 – 22

**Nyagudi Musandu Nyagudi, Independent Researcher and Security Analyst, Nairobi, KENYA**
**ORCID ID : http://orcid.org/0000-0001-5809-2209**
**Google Citations : https://scholar.google.com/citations?user=YURdCNoAAAAJ&hl=en**
**Contact : nyagudim@yahoo.com**

[18] UMC Library. (2009). AMERICAN PSYCHOLOGICAL ASSOCIATION

(APA) FORMAT (6th ed.) . University of Minnesota, Crookston

[19] Clarke, G., & Cooke, D. (1992). A Basic Course in Statistics (3rd Edition).

Kent, U.K. : Edward Arnold ( A Division of Hodder & Stoughton Ltd) -

ELBS(Educational Low-Priced Books Scheme) funded by the British

Government) with : 87 -89